\documentclass[onecolumn,12pt]{article}
\usepackage{graphicx}
\usepackage{longtable}
\usepackage{lscape}
\usepackage{color}

\newcommand\simlt{\hspace{0.3em}\raisebox{0.4ex}{$<$}\hspace{-0.75em}\raisebox{-.7ex}{$\sim$}\hspace{0.3em}}

\begin{document}
\baselineskip 14pt

\title{Distortion of Magnetic Fields in BHR 71} 
\author{Ryo Kandori$^{1}$, Motohide Tamura$^{1,2,3}$, Masao Saito$^{2}$, Kohji Tomisaka$^{2}$, \\
Tomoaki Matsumoto$^4$, Ryo Tazaki$^{5}$, Tetsuya Nagata$^{6}$, Nobuhiko Kusakabe$^{1}$, \\
Yasushi Nakajima$^{7}$, Jungmi Kwon$^{3}$, Takahiro Nagayama$^{8}$, and Ken'ichi Tatematsu$^{2}$\\
{\small 1. Astrobiology Center of NINS, 2-21-1, Osawa, Mitaka, Tokyo 181-8588, Japan}\\
{\small 2. National Astronomical Observatory of Japan, 2-21-1 Osawa, Mitaka, Tokyo 181-8588, Japan}\\
{\small 3. Department of Astronomy, The University of Tokyo, 7-3-1, Hongo, Bunkyo-ku, Tokyo, 113-0033, Japan}\\
{\small 4. Faculty of Sustainability Studies, Hosei University, Fujimi, Chiyoda-ku, Tokyo 102-8160, Japan}\\
{\small 5. Astronomical Institute, Graduate School of Science, Tohoku University,}\\
{\small 6-3 Aramaki, Aoba-ku, Sendai 980-8578, Japan}\\
{\small 6. Kyoto University, Kitashirakawa-Oiwake-cho, Sakyo-ku, Kyoto 606-8502, Japan}\\
{\small 7. Hitotsubashi University, 2-1 Naka, Kunitachi, Tokyo 186-8601, Japan}\\
{\small 8. Kagoshima University, 1-21-35 Korimoto, Kagoshima 890-0065, Japan}\\
{\small e-mail: r.kandori@nao.ac.jp}}
\maketitle

\abstract{The magnetic field structure of a star-forming Bok globule BHR 71 was determined based on near-infrared polarimetric observations of background stars. The magnetic field in BHR 71 was mapped from 25 stars. By using a simple 2D parabolic function, the plane-of-sky magnetic axis of the core was found to be $\theta_{\rm mag} = 125^{\circ} \pm 11^{\circ}$. The plane-of-sky mean magnetic field strength of BHR 71 was found to be $B_{\rm pos} = 8.8 - 15.0$ $\mu$G, indicating that the BHR 71 core is magnetically supercritical with $\lambda = 1.44 - 2.43$. Taking into account the effect of thermal/turbulent pressure and the plane-of-sky magnetic field component, the critical mass of BHR 71 was $M_{\rm cr} = 14.5-18.7$ M$_{\odot}$, which is consistent with the observed core mass of $M_{\rm core} \approx 14.7$ M$_{\odot}$ (Yang et al. 2017). We conclude that BHR 71 is in a condition close to a kinematically critical state, and the magnetic field direction lies close to the plane of sky. Since BHR 71 is a star-forming core, a significantly subcritical condition (i.e., the magnetic field direction deviating from the plane of sky) is unlikely, and collapsed from a condition close to a kinematically critical state. There are two possible scenarios to explain the curved magnetic fields of BHR 71, one is an hourglass-like field structure due to mass accumulation and the other is the Inoue \& Fukui (2013) mechanism, which proposes the interaction of the core with a shock wave to create curved magnetic fields wrapping around the core. 
}
\vspace*{0.3 cm}

\section{Introduction}
Determining the magnetic fields in and around dense molecular cloud cores is important for many fields. For example, knowing the details of magnetic support against self-gravity can allow an assessment of the kinematical stability of dense cores, which is closely related to the initial physical conditions of star formation (e.g., Crutcher 2012; Kandori et al. 2017a, hereafter Paper I). If the line-of-sight magnetic inclination angle ($\gamma_{\rm mag}$) is known, the total magnetic field strength can be obtained, allowing the magnetic criticality and kinematical stability of each core to be evaluated precisely. If a dense core is associated with hourglass-shaped magnetic fields, specific polarization patterns can appear because of the depolarization effect of the inclined distorted magnetic field structure, and $\gamma_{\rm mag}$ can be estimated through a simple model fitting (Kandori et al. 2017b, hereafter Paper II; see also Kataoka et al. 2012; Kandori et al. 2020a, Paper VI). To date, four dense cores with hourglass-shaped magnetic field structures have been identified (FeSt 1-457: Paper I, Barnard 68: Kandori et al. 2019, Barnard 335: Kandori et al. 2020c, CB81 in Pipe Nebula: Kandori et al. 2020b). One core has been explained using the Inoue--Fukui mechanism (Inoue \& Fukui 2013), showing a bending magnetic field structure (Corona Australis SL42 core: Kandori et al. 2020d). 
\par
Determining the depolarization pattern as well as $\gamma_{\rm mag}$ is also important to correct the correlation between polarization and extinction (Kandori et al. 2018a, hereafter Paper III, see also Paper VI) and to compare the polarization data taken in different wavelengths (Kandori et al. 2018c, hereafter Paper V). Using the Davis--Chandrasekhar--Fermi technique in a modified form can help determine the radial distribution of the mass-to-magnetic flux ratio in a core (Kandori et al. 2018b, hereafter Paper IV). 
\par
Comparing observed hourglass-shaped field data and a theoretical hourglass field model assuming flux freezing (Paper VI; see also Mestel 1966; Ewertowski \& Basu 2013; Myers et al. 2018) can help determine the core formation parameters, namely the initial density $\rho_0$, initial radius $R_0$ (i.e., core formation radius), and initial magnetic field strength $B_0$. These quantities are essential for discussing the formation and evolution of molecular clouds and cores. 
\par
As part of our magnetic field survey of dense cores, we investigated the BHR 71 core (Bourke et al. 1995a) located near the Coalsack region in the southern sky using near-infrared (NIR) polarimetry. BHR 71 is a Bok globule (Bok \& Reilly 1947) and has several designations, such as Sa 136 (Sandqvist 1977), which is an extension of the catalog by Sandqvist \& Lindroos (1976), and DC 297.7-2.8 (Hartley et al. 1986). In this paper, we use the name BHR 71 following Bourke et al. (1995a). 
\par
The distance to the BHR 71 core is often assumed to be the same as the distance to the Coalsack cloud, which is located in the vicinity of BHR 71. The distance to the Coalsack is generally taken to be 200 pc (Seidensticker \& Schmidt-Kaler 1989; Straizys et al. 1994), but it may be as close as $150 \pm 30$ pc (Corradi et al. 1997). Voirin et al. (2018) estimated a distance of $176 \pm 7$ pc based on the {\it Gaia} data, and this is close to the values measured by Rodgers (1960) and Franco (1989) of $174 \pm 18$ pc and $180 \pm 26$ pc, respectively. We use Voirin's value of 176 pc in this paper. 
\par
NH$_3$ observations were made by Bourke et al. (1995b), and the kinematic temperature of BHR 71 was estimated to be 13 K. NH$_3$ (1,1) mapping observations with the Parkes 64-m radio telescope gave a full width at half maximum (FWHM) line width of $\sim 0.5$ km s$^{-1}$ over the core (Tobin et al. 2019). The turbulent velocity dispersion $\sigma_{\rm turb}$ of the core is thus 0.20 km s$^{-1}$. 
\par
The radius and mass of BHR 71 at a distance of 176 pc based on the {\it Herschel} spectrophotometry data were 0.28 pc and 14.7 M$_{\odot}$, respectively (Yang et al. 2017). However, these quantities depend on the (molecular) probe used in the observations. For example, at the same distance, the radius and mass of the core measured in NH$_3$ (1,1) were 0.07 pc and 2.3 M$_{\odot}$, respectively (Bourke et al. 1997). In this study, we use the value provided by Yang et al. (2017). 
\par
Targon et al. (2011) conducted optical ($R_{\rm C}$ band) polarimetry around the BHR 71 core. The obtained polarization angle (measured from north to east) was $102^{\circ}$, indicating that the direction of the global magnetic fields is roughly parallel to the elongation axis of BHR 71 ($\sim 130^{\circ}$, see Figure 1, showing the elongation of the opaque region of BHR 71). 
\par
BHR 71 is a binary protostellar core, consisting of IRS1 and IRS2 separated by $\sim 15''$ or $\sim 2600$ au at a distance of 176 pc (Bourke et al. 1997; Bourke 2001). IRS1 is a Class 0 protostar (Bourke 2001; Green et al. 2013), but IRS2 may be more evolved (Bourke 2001). These protostars have misaligned outflows (Bourke 2001; Parise et al. 2006). Furthermore, Tobin et al. (2019) reported that the rotation of these protostars measured in C$^{18}$O ($J=1-0$) on scales of $<1000$ au around each source is in the opposite direction. These results may be evidence these protostars formed by turbulent fragmentation (Tobin et al. 2019), and is consistent with a theoretical simulation of the formation of binary/multiple sources in turbulent environments (Offner et al. 2016). 
\par
In the present study, wide-field background star polarimetry at NIR wavelengths was conducted for BHR 71. The plane-of-sky magnetic field structure was revealed using stars in and around the core. The plane-of-sky magnetic field strength was estimated based on the Davis--Chandrasekhar--Fermi method (Davis 1951; Chandrasekhar \& Fermi 1953). Using the resulting magnetic field information, the kinematical stability and the origin of the distorted magnetic field structure in BHR 71 are discussed. 

\section{Observations}
The observations of BHR 71 were conducted using the $JHK_s$-simultaneous near-infrared polarimeter SIRPOL (Kandori et al., 2006, see also Nagayama et al. 2003 for camera) on the IRSF 1.4-m telescope in the South African Astronomical Observatory (SAAO). IRSF/SIRPOL provides a large field of view ($7.7' \times 7.7'$ with a scale of 0.45$''$ pixel$^{-1}$), so that nearby dense cloud cores can be covered with a single pointing. Since SIRPOL is a single beam polarimeter, sky changes during the exposures at different half-waveplate angles. Typical uncertainty of polarization degree due to sky variation in a photometric night is about 0.3\%. The instrumental polarization over the field of view is less than 0.3\% (Kandori et al. 2020b). The accuracy of the zero point angle of the polarimeter is less than $3^{\circ}$. A polarimetric standard star, RCrA\#88 ($P_H = 2.73\% \pm 0.07\%$, $\theta_H = 92^{\circ} \pm 1^{\circ}$, Whittet et al. 1992), was observed on July 13, 2017, and we obtained the results ($P_H = 2.82\% \pm 0.07\%$ and $\theta_H = 91.9^{\circ} \pm 0.9^{\circ}$) which are consistent with the values in the literature. 
\par
We observed BHR 71 on the nights of June 13-16 in 2017. Exposures of 10s were performed at four half-waveplate angles (in the sequence of $0^{\circ}$, $45^{\circ}$, $22.5^{\circ}$, and $67.5^{\circ}$) at ten dithered positions (one set). The total integration time was 4500s (30 sets) per waveplate angle. Sky frames were observed before or after the observations of object frames. 
\par
Data were reduced using the Interactive Data Language (IDL) software. The reduction procedures include dark subtraction, flat-field correction, median sky subtraction, and frame combining after registration (see, e.g., Kandori et al. 2007). 
\par
Point sources having a peak intensities greater than 10-sigma above local sky background were catalogued on the Stokes $I$ images. Aperture polarimetry was performed for these sources on the images take at four waveplate angles ($I_{0^{\circ}}$, $I_{45^{\circ}}$, $I_{22.5^{\circ}}$, and $I_{67.5^{\circ}}$). Sources with photometric error greater than 0.1 mag were removed from the list, and 3442 sources were detected in the $H$ band. The aperture radius was the same as the FWHM of stars (3.5 pixel). The sky radius and the width of sky annulus were set to 10 and 5 pixels, respectively. The limiting magnitudes were 18.0 mag in the $H$ band. The relatively small aperture size was used to suppress the flux contamination from neighboring bright stars. A technique suitable for the photometry in a crowded field is the point spread function (psf) fitting photometry. We did not employ this technique, because the different goodness of fit for each star on different waveplate angle images can cause systematic errors in polarimetric measurements. 
\par
The Stokes parameters for each star were derived with the relationship $I = (I_{0^{\circ}}+I_{45^{\circ}}+I_{22.5^{\circ}}+I_{67.5^{\circ}})/2$, $Q = I_{0^{\circ}} - I_{45^{\circ}}$, and $U = I_{22.5^{\circ}} - I_{67.5^{\circ}}$. The polarization degree ($P$) and polarization angle ($\theta$) were determined from the equations $P = \sqrt{Q^2 + U^2}/I$ and $\theta = 0.5 {\rm atan} (U/Q)$, respectively. Because $P$ is a positive quantity, the resulting $P$ values tend to be overestimated especially for low S/N sources. We corrected for the bias in $P$ measurements using the equation $P_{\rm db} = \sqrt{P^2 - \delta P^2}$ (Wardle \& Kronberg 1974). In this paper, we focus on the results in the $H$ band, in which dust extinction is less severe in the $J$ band and polarization efficiency is greater than in the $K_s$ band.

\section{Results and Discussion}
\subsection{Distortion of Magnetic Fields}
Figure 1 shows the observed polarization vectors (yellow vectors) on the Stokes $I$ image in the $H$ band. To draw the vectors, we used 140 stars with $P_H / \delta P_H \ge 4$. The BHR 71 core appears as a dark obscuration around the center of the image, and the cone-shaped illumination of the outflow cavity wall by the protostar IRS1 can be seen. The shape of the dust obscuration of BHR 71 is elongated with the long axis toward the direction of $\sim 130^{\circ}$. There are many relatively small polarization vectors around the south-east corner and the south-west corner of the image, flowing from north-west to south-east. Reflecting this, the histogram of $\theta_H$ for all sources has a peak fitted with a Gaussian function with a peak angle of $111^{\circ} \pm 4^{\circ}$ (Figure 2), which is consistent with the optical polarimetry results (Targon et al. 2011). 
\par
The most striking feature in Figure 1 is the steeply curved polarization vector pattern mostly located in the northern side of the BHR 71 core. Strong polarization was associated with the curved field structure. If we isolate the strong polarization vectors in the field of view, only the observation data associated with the curved field component remain. Furthermore, they are all inside the radius of BHR 71. These vectors appear to be different from the $\theta_H = 111^{\circ}$--component, and are likely associated with the BHR 71 core. 
\par
To determine the basic physical properties of the BHR 71 core, we used the FIR data taken by the {\it Herschel} satellite. Sadavoy et al. (2018) provided an intensity-corrected {\it Herschel} map of BHR 71, and a map at an optical depth of 353 GHz ($\tau_{353 {\rm GHz}}$) was included in their products. The map has $28 \times 28$ pixels with a pixel scale of $14''$ pixel$^{-1}$ and resolution of $36.3''$. The molecular hydrogen column density can be obtained from 
\begin{equation}
N({\rm H}_2) = \frac{\tau_{353 {\rm GHz}}}{\kappa_{353 {\rm GHz}}\cdot \mu_{\rm H_2}\cdot m_{\rm H}}, 
\end{equation}
where $\kappa_{353 {\rm GHz}} = 0.0182$ cm$^2$ g$^{-1}$ (Kauffmann et al. 2008) is the dust opacity or absorption cross section per unit mass of gas at 353 GHz, $\mu_{\rm H_2} = 2.8$ is the molecular weight per hydrogen molecule (assuming cosmic abundance ratios), and $m_{\rm H}$ is the hydrogen atom mass. Using the above equation we converted the $\tau_{353 {\rm GHz}}$ map to a molecular hydrogen column density map (see background image in Figures 9 and 12). The centroid center of the core was determined to be (R.A., Decl.) = ($12^{\rm h}01^{\rm m}36.13^{\rm s}, -65^{\circ}08^{'}46.7^{''}$, J2000). The elongation of the core was determined based on an ellipse fit to the region $N({\rm H}_2) > 10^{22}$ cm$^{-2}$, giving $\approx 133^{\circ}$ as the direction of its major axis. 
\par
To analyze the NIR polarimetry data further, we need to separate the global $\theta_H = 111^{\circ}$--component (ambient component) and the polarization component arisen from the BHR 71 core. While one simple way is to just ignore the weak polarization vectors and to use strong polarization vectors, we decided to use the method of ambient vector field subtraction following the methods described in Paper I. Note that the former method can provide results that are consistent with the latter method. 
\par
As shown in Figure 2, the ambient component of the polarization vectors is relatively well aligned. Spatial linear plane fitting of the component was thus conducted. To exclude the positions passing through the dense core region, the stars located within 300 pixels ($135'' \approx 24000$ au) of the core center were not used in the fitting. The distributions of the $Q/I$ and $U/I$ values were independently modeled as $f(x,y) = A + Bx + Cy$, where $x$ and $y$ are the pixel coordinates and $A$, $B$, and $C$ are the fitting parameters. Figures 3 and 4 show the histograms of $P_H$ or $\theta_H$ of stars before (solid line) and after (dotted-dashed line) the subtraction. The results did not change the polarization vectors dramatically. However, after subtraction, the number of low $P_H$ vectors increased in Figure 3, and $\theta_H$ became roughly randomly distributed in Figure 4. Therefore, the subtraction analysis works somewhat satisfactory. Figure 5 shows the distribution of estimated ambient polarization vectors. Most of the vectors are as strong as $P_H \sim 1\%$, distributed mainly in the southern half of the image. Figure 6 shows the polarization vectors after the subtraction of ambient vectors. The small polarization vectors distributed around the south-east corner and the south-west corner of the image disappeared, and the curved polarization vectors associated with the dense part of the core were reasonably well extracted. 
\par
Figure 7 shows the relationship between $H - K_s$ color (i.e., $A_V$) and polarization degree toward the background stars with $P_H$ taken after subtraction of the ambient component. Stars with $P_H / \delta P_H \ge 4$ were plotted. Though the scatter in the diagram is not small, the distribution of data points is positive, showing that the observations trace the dust polarization (i.e., magnetic field direction) inside a cold and dense environment. The slope of the relationship $P_H / (H-K_s)$ was found to be $2.52 \pm 0.04$ \% mag$^{-1}$. 

\subsection{Parabolic Model}
Since the global direction of the magnetic fields (Figure 2) and the curved field associated with BHR 71 (Figure 6) are clear, we tried to fit the magnetic field structure of the core using a simple parabolic function, $y = g + gCx^2$, where $g$ specifies each magnetic field line and $C$ determines the curvature of the parabola. We used the function in a $90^{\circ}$-rotated form, so that the $x$ axis is toward $0^{\circ}$ in position angle. We calculated the parabolic function for various $C$, various spatial positions, and various rotation angles to find the parameter that minimizes 
\begin{equation}
\chi^2 = \sum^n_{i=1}\frac{(\theta_{{\rm obs,}i} - \theta_{\rm model}(x_i,y_i))^2}{\delta \theta^2_i},
\end{equation}
where $n$ is the number of stars, $x$ and $y$ are the coordinates of the $i$-th star, $\theta_{\rm obs}$ and $\theta_{\rm model}$ are the polarization angles from the observations and the model, and $\delta \theta_i$ is the observational error. We used 25 stars with $P_H / \delta P_H \ge 4$ and $P_H \ge 1.5\%$. The latter threshold was set so as to exclude the stars affected by the ambient subtraction analysis. In Figure 3, most of the ambient sources show $P_H \simlt 1.5\%$. 
\par
Figure 8 shows the result of the parabolic fit (white lines). Figure 9 is the same as Figure 8 but the column density map based on the {\it Herschel} data was used as a background image. The prominent curved magnetic fields located in the north-east part of the image were relatively well fitted with the parabolic function, whereas there are small number of vectors in the south-west part of the image. In Figures 8 and 9, the red plus signs show the center of the core and the blue plus signs show the center of the parabolic fields. The coordinate of the magnetic center is (R.A., Decl.) = ($12^{\rm h}01^{\rm m}34.50^{\rm s}, -65^{\circ}09^{'}01.4^{''}$, J2000). 
\par
The direction of the magnetic axis ($\theta_{\rm mag}$) and the magnetic curvature ($C$) were obtained to be $\theta_{\rm mag} = 125^{\circ} \pm 11^{\circ}$ and $C = 1.0(\pm 0.2) \times 10^{-3}$ arcsec$^{-2}$. The parabolic fit is good for the curved magnetic field part, while the rest of the vectors tend to deviate from the parabolic line. If we only use the stars in the curved field part in the north-east half of the image, the standard deviation of the residual angles ($\theta_{\rm res} = \theta_{\rm obs} - \theta_{\rm fit}$) is $\delta \theta_{\rm res, curve} = 15.1^{\circ}$. If we use all the polarization vectors, the value increases to $\delta \theta_{\rm res, all} = 24.9^{\circ}$. Figure 10 shows a histogram of $\theta_{\rm res}$ for all the stars. We subtracted the effect from the observational error using $\delta \theta_{\rm int} = (\delta \theta_{\rm res} - \delta \theta_{\rm err})^{1/2}$, where $\delta \theta_{\rm err}$ is the observational error, and obtained $\delta \theta_{\rm int, curve} = 14.6^{\circ}$ and $\delta \theta_{\rm int, all} = 24.6^{\circ}$. Note that both values are less than the values for the uniform field case ($\delta \theta_{\rm res, ini} = 22.7^{\circ}$ and $29.6^{\circ}$, corresponding to the vectors in the curved field regions and all the vectors). 
\par
The strength of the plane-of-sky magnetic fields ($B_{\rm pos}$) can be estimated from the Davis--Chandrasekhar--Fermi equation (Davis 1951; Chandrasekhar \& Fermi 1953)
\begin{equation}
B_{\rm pos} = C_{\rm corr} \sqrt{4 \pi \rho} \frac{\sigma_{\rm turb}}{\delta \theta_{\rm int}}, 
\end{equation}
where $\rho$ is the mean density ($1.2 \times 10^{-20}$ g cm$^{-3}$ from Yang et al. 2017), $\sigma_{\rm turb}$ is the turbulent velocity dispersion (0.20 km s$^{-1}$ from Tobin et al. 2019), $\delta \theta_{\rm int}$ is $14.6^{\circ} - 24.6^{\circ}$ ($0.255 - 0.429$ radian, this study), and $C_{\rm corr}=0.5$ is the correction factor from theory (Ostriker et al. 2001, see also Padoan et al. 2001; Heitsch et al. 2001; Heitsch 2005; Matsumoto et al. 2006). From the above equation, we obtained a relatively weak magnetic field strength of $B_{\rm pos} = 8.8 - 15.0$ $\mu$G. Note that the estimated field strength is the averaged value for the whole core. 

\subsection{Magnetic Properties of the Core and the Origin of the Curved Magnetic Fields}
In this section, we first discuss the plane-of-sky magnetic properties of BHR 71, and then discuss the origin of the curved magnetic fields. Note that we did not conduct a three-dimensional (3D) analysis of the curved field, because the number of stars is not sufficient. 
\par
The magnetic support of BHR 71 against gravity can be investigated using the parameter $\lambda = (M/\Phi)_{\rm obs} / (M/\Phi)_{\rm critical}$, which represents the ratio of the observed mass-to-magnetic flux to a critical value, $(2\pi G^{1/2})^{-1}$, suggested by theory (Mestel \& Spitzer 1956; Nakano \& Nakamura 1978). We determined a value of $\lambda = 1.44 - 2.43$, indicating a magnetically supercritical state. The magnetic critical mass of the core, $M_{\rm mag} = 6.0 - 10.2$ M$_{\odot}$, is lower than the observed core mass of $M_{\rm core} \approx 14.7$ M$_{\odot}$ (Yang et al. 2017). Note that this does not necessarily imply a dynamical collapse of BHR 71, because there are additional thermal and turbulent supports. 
\par
The critical mass of BHR 71, taking into account both magnetic and thermal/turbulent support effects, can be written as $M_{\rm cr} \simeq M_{\rm mag} + M_{\rm BE}$ (Tomisaka, Ikeuchi, \& Nakamura 1988; McKee 1989), where $M_{\rm BE}$ is the Bonnor--Ebert mass (Ebert 1955; Bonnor 1956). We obtained $M_{\rm cr} = 14.5 - 18.7$ M$_{\odot}$ with $M_{\rm BE}$ 8.5 M$_{\odot}$. The Bonnor--Ebert mass was estimated using a kinematic temperature $T_{\rm kin}$ of 13 K (Bourke et al. 1995b), a turbulent velocity dispersion of 0.2 km s$^{-1}$ (Tobin et al. 2019), and an assumed external pressure of $6 \times 10^4$ K cm$^{-3}$. The assumed external pressure is equivalent to $T_{\rm eff} \times \rho$, where $\rho$ is the mean density of the core and $T_{\rm eff}$ is the sum of the kinematic temperature $T_{\rm kin}$ and the turbulence equivalent temperature $T_{\rm turb} = \mu_{\rm p} \sigma_{\rm turb}^2 / k = 10.9$ K. The mean molecular weight per free particle $\mu_{\rm p}$ is set to 2.33 and $k$ is the Boltzmann constant. The obtained critical mass of $M_{\rm cr} = 14.5 - 18.7$ M$_{\odot}$ is not far from the core mass $M_{\rm core} \approx 14.7$ M$_{\odot}$, suggesting a nearly critical state. 
\par
Since BHR 71 is a star-forming core, a stable subcritical state (i.e., large $M_{\rm cr}$) for this core is unlikely. We found that $M_{\rm cr} \sim M_{\rm core}$, based on a plane-of-sky magnetic field strength $B_{\rm pos}$. This means that the line-of-sight magnetic field inclination angle cannot deviate from the plane of sky. If it did deviate, we would observe a large total field strength, which would lead to a large $M_{\rm cr}$. Thus, we conclude that the magnetic fields pervading BHR 71 lie near the sky plane, and the core started its collapse from a state near the kinematically critical condition. 
\par
Finally we propose a possible scenario for the origin of the curved magnetic fields in BHR 71. As shown in Figures 11 and 12, the magnetic fields pervading BHR 71 can be explained by a single curved field structure. The field lines (white lines) in Figures 11 and 12 are the same as those in Figures 8 and 9, but the southern field components are removed. The best model to explain this is the Inoue \& Fukui (2013) mechanism, describing a shock wave propagating from the south-west in the direction of the position angle of $35^{\circ}$ (perpendicular to $\theta_{\rm mag}$), sweeping the magnetic fields, such that the magnetic fields wrap around the core to create the curved magnetic field structure. Although we cannot specify the origin of the shock, BHR 71 is located in the Lower Centaurus-Crux association as a subgroup of the Scorpius--Centaurus association. The curved magnetic fields may be a remnant of the interaction of the core with the past shock wave. Note that the hourglass shape in Figures 8 and 9 is still possible, and thus we have two scenarios to explain the origin of the curved magnetic fields in BHR 71. To determine which of these the scenarios applies, large-scale radio molecular line observations, particularly in $^{12}$CO and $^{13}$CO ($J=1-0$), are necessary because, as shown by Arzoumanian et al. (2018), the Inoue \& Fukui (2013) mechanism creates specific features in the position-velocity diagram. 

\section{Summary and Conclusion}
The magnetic field structure of a star-forming Bok globule BHR 71 was determined based on NIR polarimetric observations of background stars to measure dichroically polarized light produced by magnetically aligned grains. The magnetic fields for BHR 71 were mapped using 25 stars, and curved magnetic fields were identified. Based on simple 2D modeling using a parabolic function, the magnetic axis of the core on the plane of sky was determined to be $\theta_{\rm mag} = 125^{\circ} \pm 11^{\circ}$, which is consistent with the global magnetic field direction measured with optical polarimetry and the core's ambient polarization direction measured in our observations as $111^{\circ} \pm 4^{\circ}$. The plane-of-sky mean magnetic field strength of BHR 71 was found to be $B_{\rm pos} = 8.8 - 15.0$ $\mu$G, indicating that the BHR 71 core is magnetically supercritical with $\lambda = 1.44 - 2.43$. Taking into account the effect of thermal/turbulent pressure and the plane-of-sky magnetic field component, the critical mass of BHR 71 was $M_{\rm cr} = M_{\rm mag} + M_{\rm BE} = 14.5-18.7$ M$_{\odot}$, which is consistent with the observed core mass of $M_{\rm core} \approx 14.7$ M$_{\odot}$. The magnetic critical mass was calculated to be $M_{\rm mag} = 6.0 - 10.2$ M$_{\odot}$ and the Bonnor--Ebert mass is $M_{\rm BE} = 8.5$ M$_{\odot}$. We conclude that BHR 71 is in a condition close to a kinematically critical state, and the magnetic field direction lies close to the plane of sky. Since BHR 71 is a star-forming core, a significantly subcritical condition (i.e., the magnetic field direction deviating from the plane of sky) is unlikely. BHR 71 most likely started to collapse from a condition close to a kinematically critical state. There are two possible scenarios to explain the curved magnetic fields of BHR 71: one is the hourglass-like field structure and the other is the Inoue \& Fukui (2013) mechanism, which proposes the interaction of the core with a shock wave to create curved magnetic fields wrapping around the core. To determine which scenario applies, large-scale radio molecular line observations are necessary, because the Inoue \& Fukui (2013) mechanism creates specific features in the position-velocity diagram.

\subsection*{Acknowledgement}
The contributions of Tsuyoshi Inoue are gratefully acknowledged. We are also grateful to the staff of SAAO for their help during the observations. We with to thank Tetsuo Nishino, Chie Nagashima, and Noboru Ebizuka for their support in the development of SIRPOL, its calibration, and stable operation with the IRSF telescope. The IRSF/SIRPOL project was initiated and supported by Nagoya University, National Astronomical Observatory of Japan, and the University of Tokyo in collaboration with the South African Astronomical Observatory under the financial support of Grants-in-Aid for Scientific Research on Priority Area (A) No. 10147207 and No. 10147214, and Grants-in-Aid No. 13573001 and No. 16340061 of the Ministry of Education, Culture, Sports, Science, and Technology of Japan. MT and RK acknowledge support by the Grants-in-Aid (Nos. 16077101, 16077204, 16340061, 21740147, 26800111, and 19K03922).

\clearpage 

\begin{figure}[t]
\begin{center}
 \includegraphics[width=6.5 in]{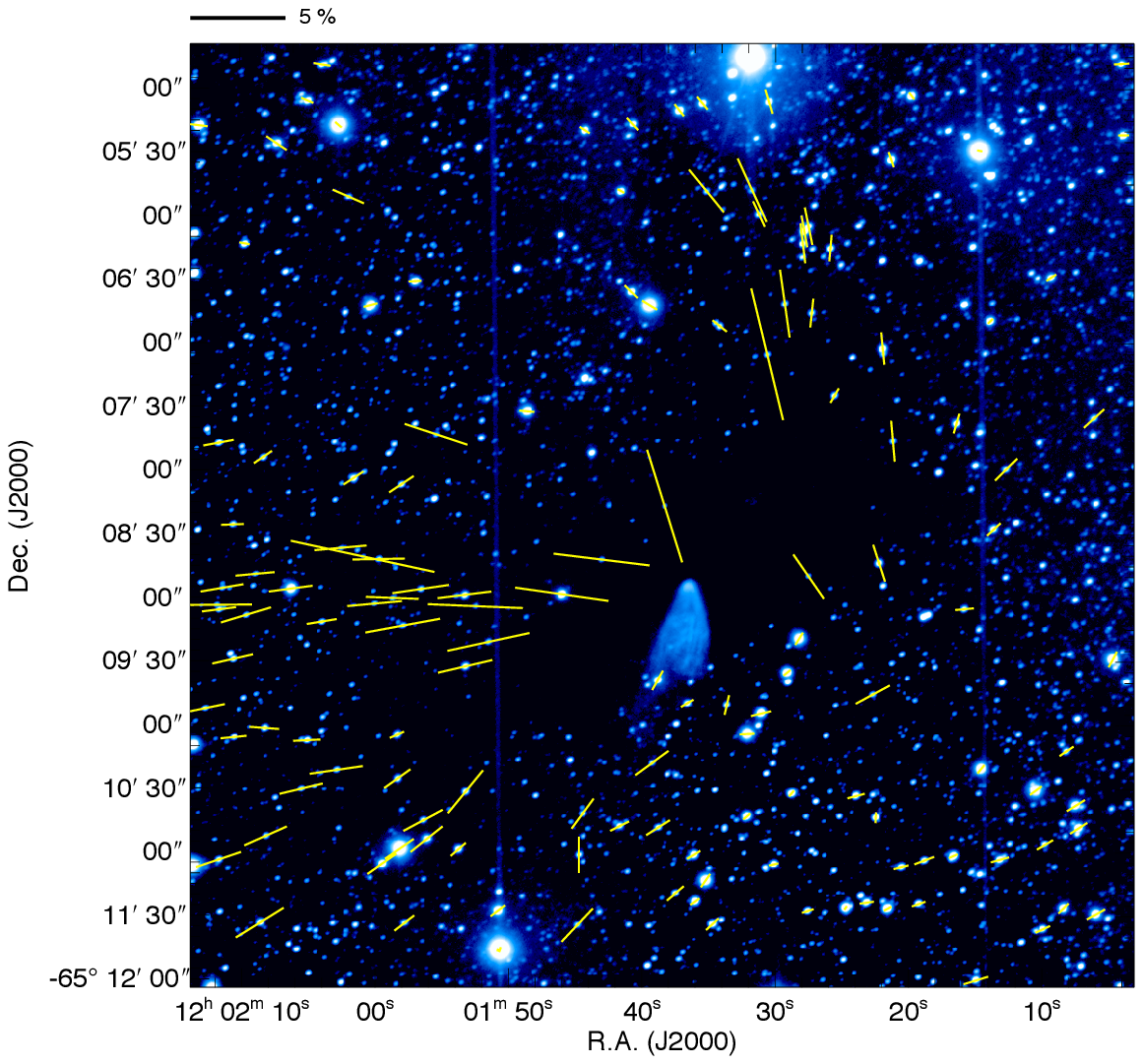}
\end{center}
 \caption{Polarization vectors of point sources superimposed on the Stokes $I$ image in the $H$ band. Stars with $P_H / \delta P_H \ge  4$ are shown. The scale bar above the image indicates 5\% polarization.}
   \label{fig}
\end{figure}

\clearpage 

\begin{figure}[t]
\begin{center}
 \includegraphics[width=6.5 in]{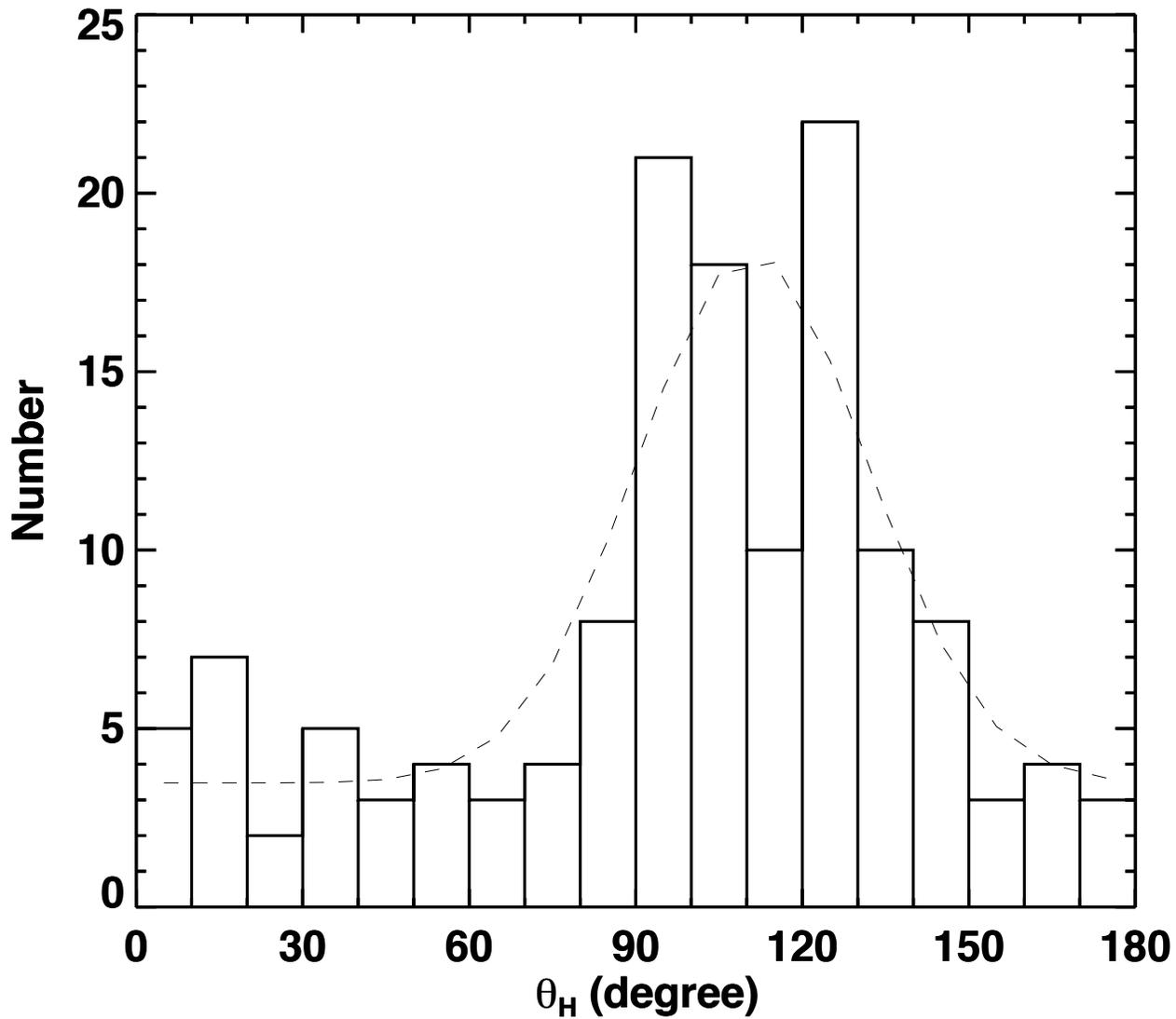}
\end{center}
 \caption{Histogram of $\theta_H$ for all the stars ($N=140$). The dashed line shows a fit using a Gaussian function.}
   \label{fig}
\end{figure}

\clearpage 

\begin{figure}[t]
\begin{center}
 \includegraphics[width=6.5 in]{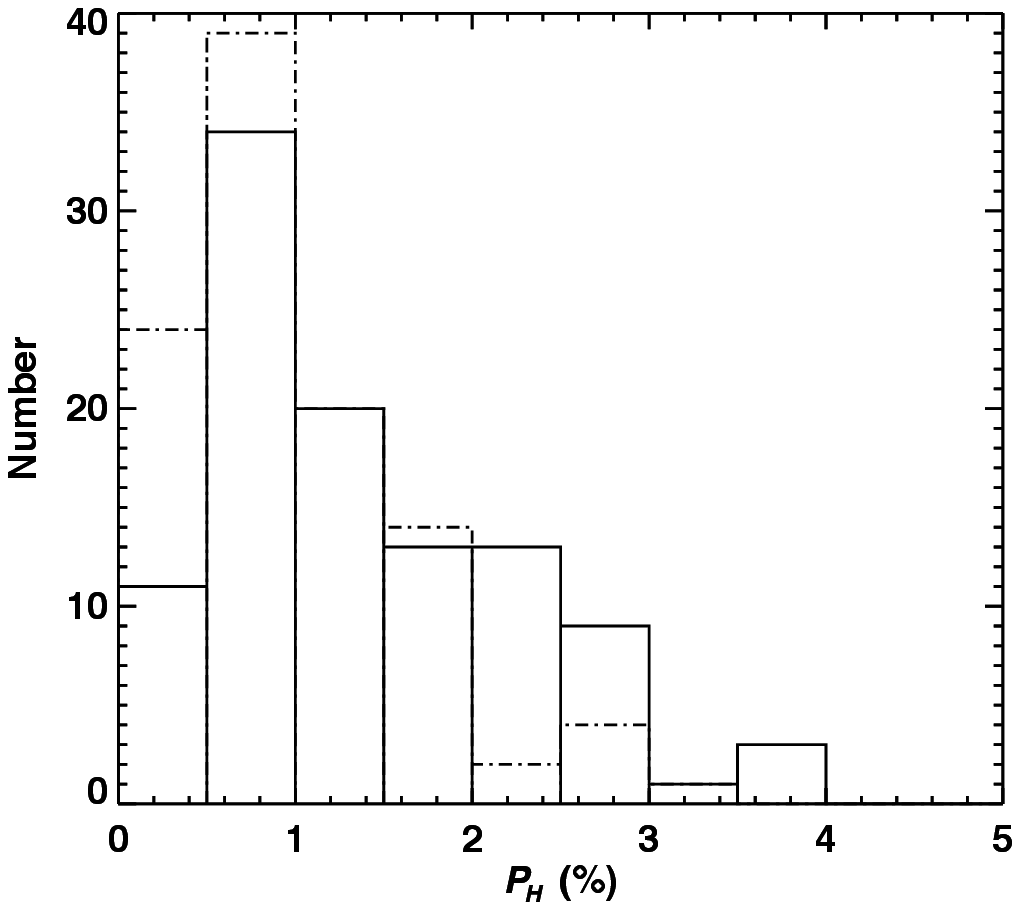}
\end{center}
 \caption{Histogram of $P_H$ for the ambient stars before (solid line) and after (dotted-dashed line) subtraction of the ambient component.}
   \label{fig}
\end{figure}

\clearpage 

\begin{figure}[t]
\begin{center}
 \includegraphics[width=6.5 in]{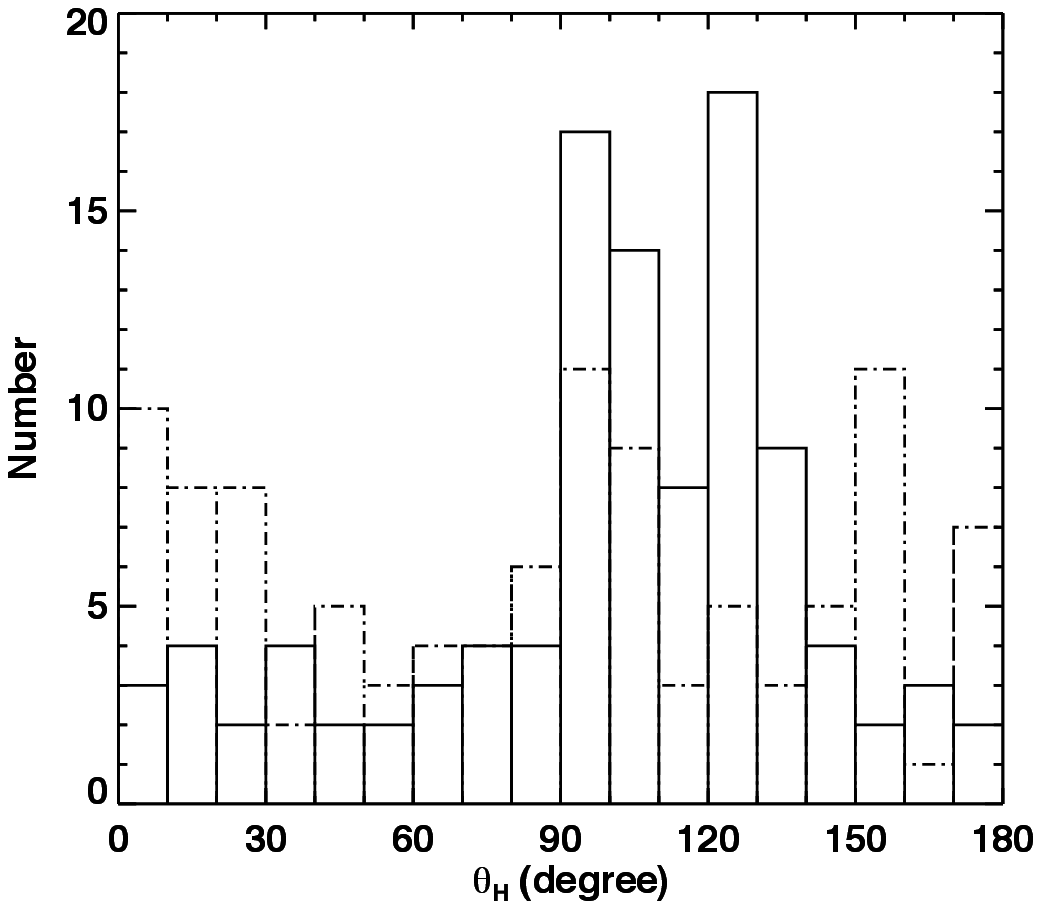}
\end{center}
 \caption{Histogram of $\theta_H$ for the ambient stars before (solid line) and after (dotted-dashed line) subtraction of the ambient component.}
   \label{fig}
\end{figure}

\clearpage 

\begin{figure}[t]
\begin{center}
 \includegraphics[width=6.5 in]{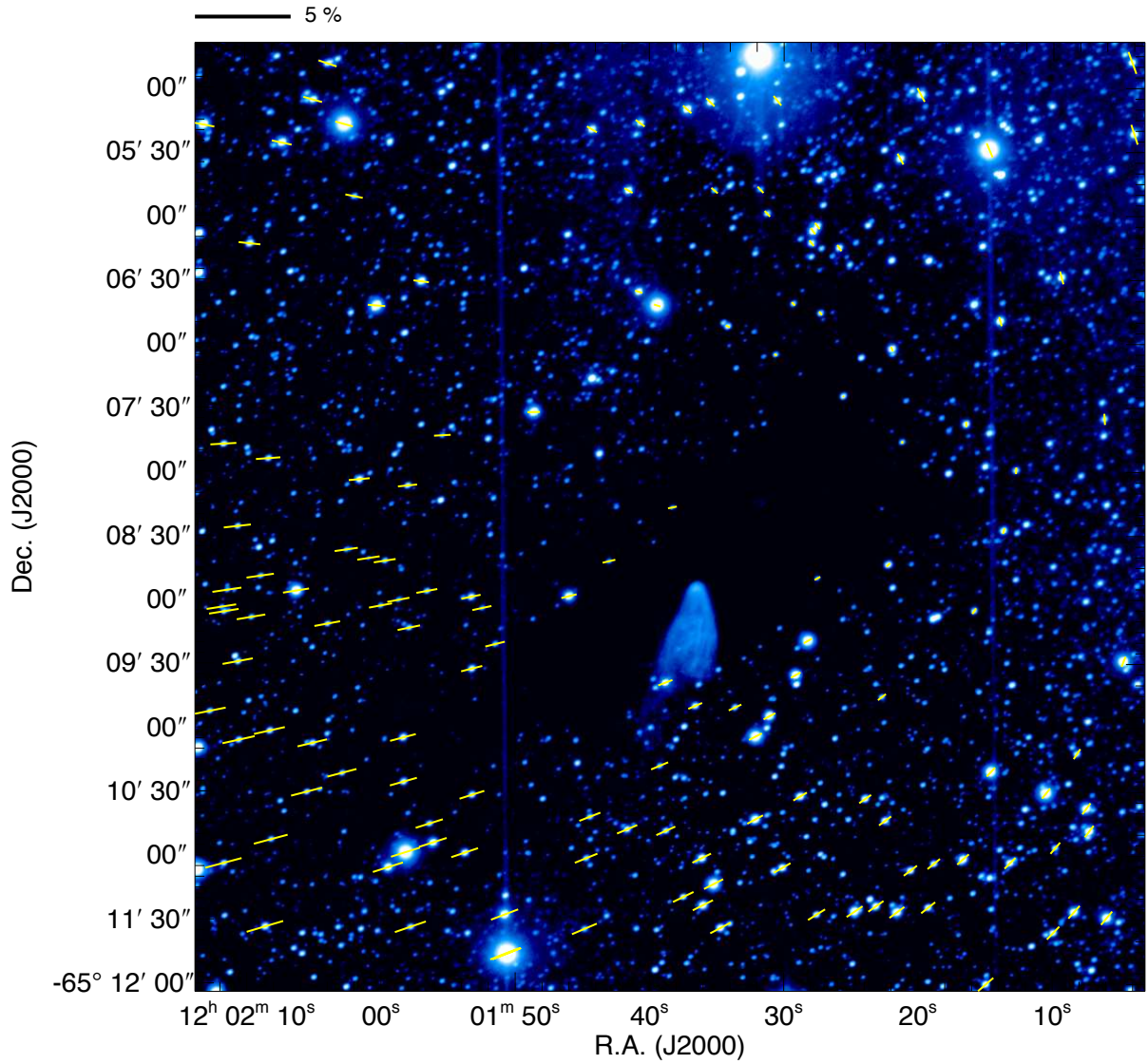}
\end{center}
 \caption{Estimated ambient polarization vectors superimposed on the Stokes $I$ image in the $H$ band. The scale bar above the image indicates 5\% polarization.}
   \label{fig}
\end{figure}

\clearpage 

\begin{figure}[t]
\begin{center}
 \includegraphics[width=6.5 in]{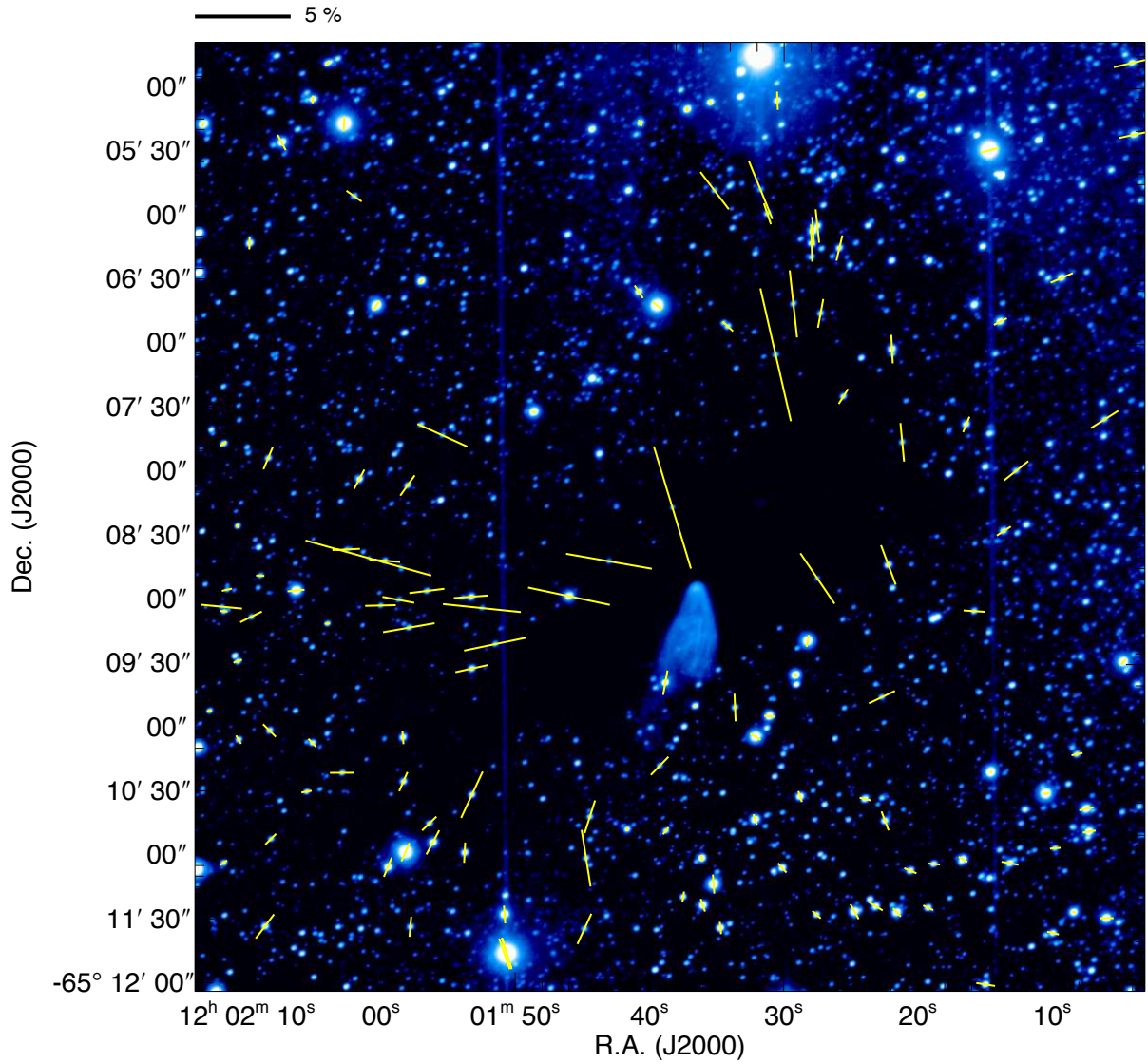}
\end{center}
 \caption{Polarization vectors after subtraction of the ambient component. The scale bar above the image indicates 5\% polarization.}
   \label{fig}
\end{figure}

\clearpage 

\begin{figure}[t]
\begin{center}
 \includegraphics[width=6.5 in]{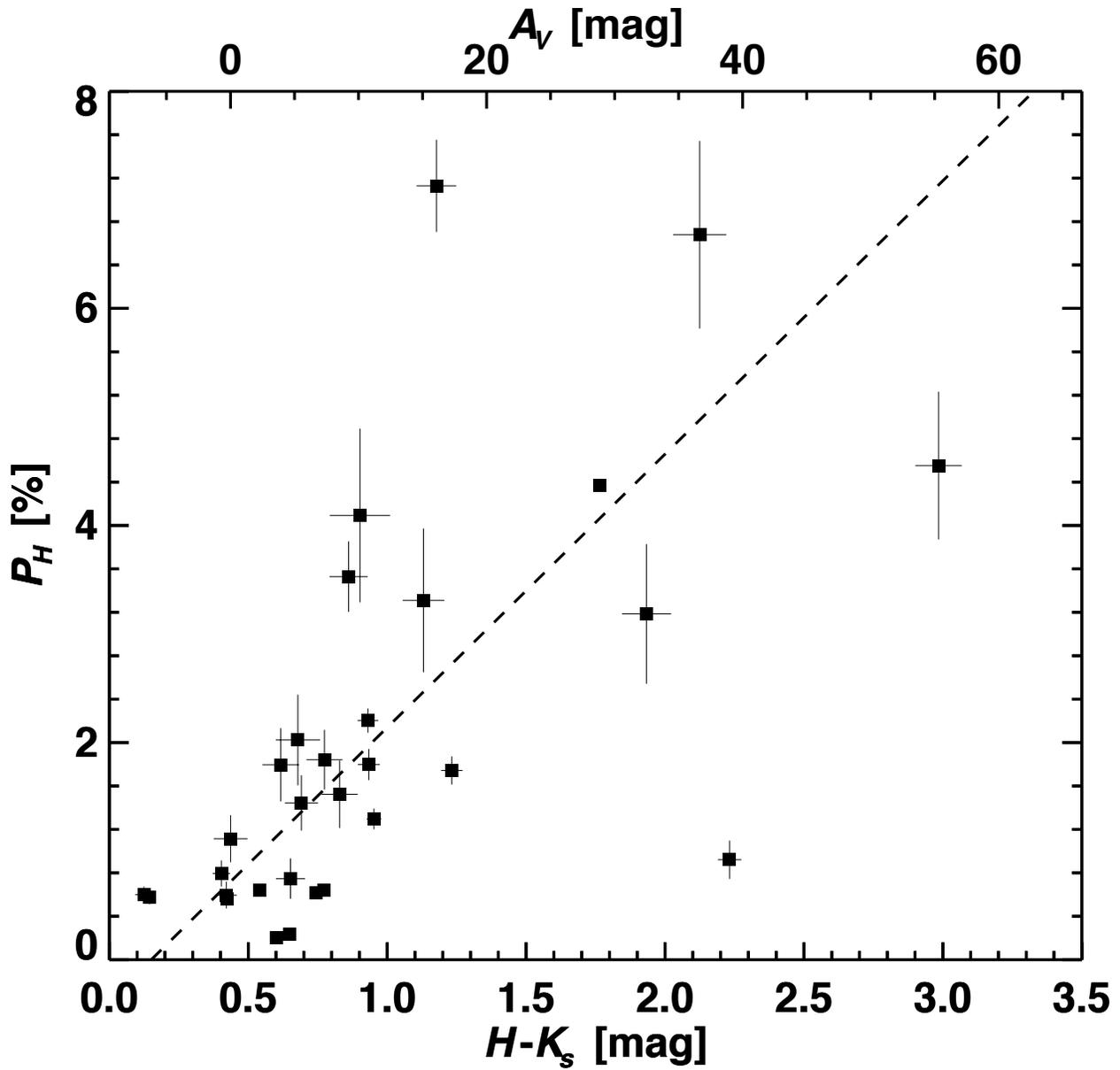}
\end{center}
 \caption{Relationship between $H-K_s$ color and polarization degree toward background stars with $P_H$ taken after the subtraction of the ambient component. Stars with $P_H / \delta P_H \ge 4$ are plotted.}
   \label{fig}
\end{figure}

\clearpage 

\begin{figure}[t]
\begin{center}
 \includegraphics[width=6.5 in]{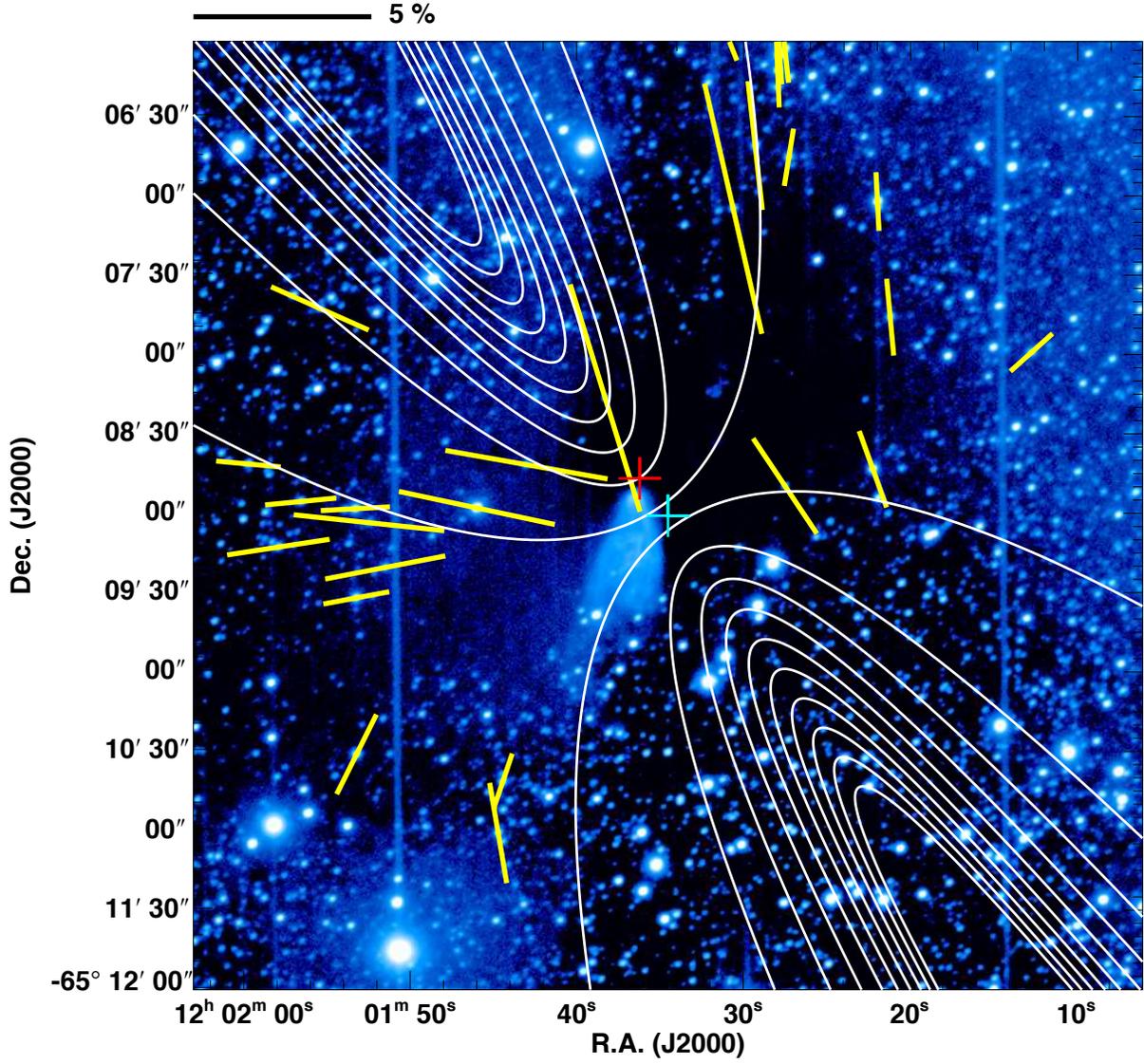}
\end{center}
 \caption{Polarization vectors after subtraction of the ambient components. Stars with $P_H / \delta P_H \ge 4$ and $P_H \ge 1.5\%$ are plotted. The white lines indicate the direction of the magnetic field inferred from parabolic fitting. The red plus sign shows the core center determined on the column density map based on {\it Herschel}. The blue plus sign shows the center of the curved magnetic fields (i.e., parabolic function). The scale bar above the image indicates 5\% polarization.}
   \label{fig}
\end{figure}

\clearpage 

\begin{figure}[t]
\begin{center}
 \includegraphics[width=6.5 in]{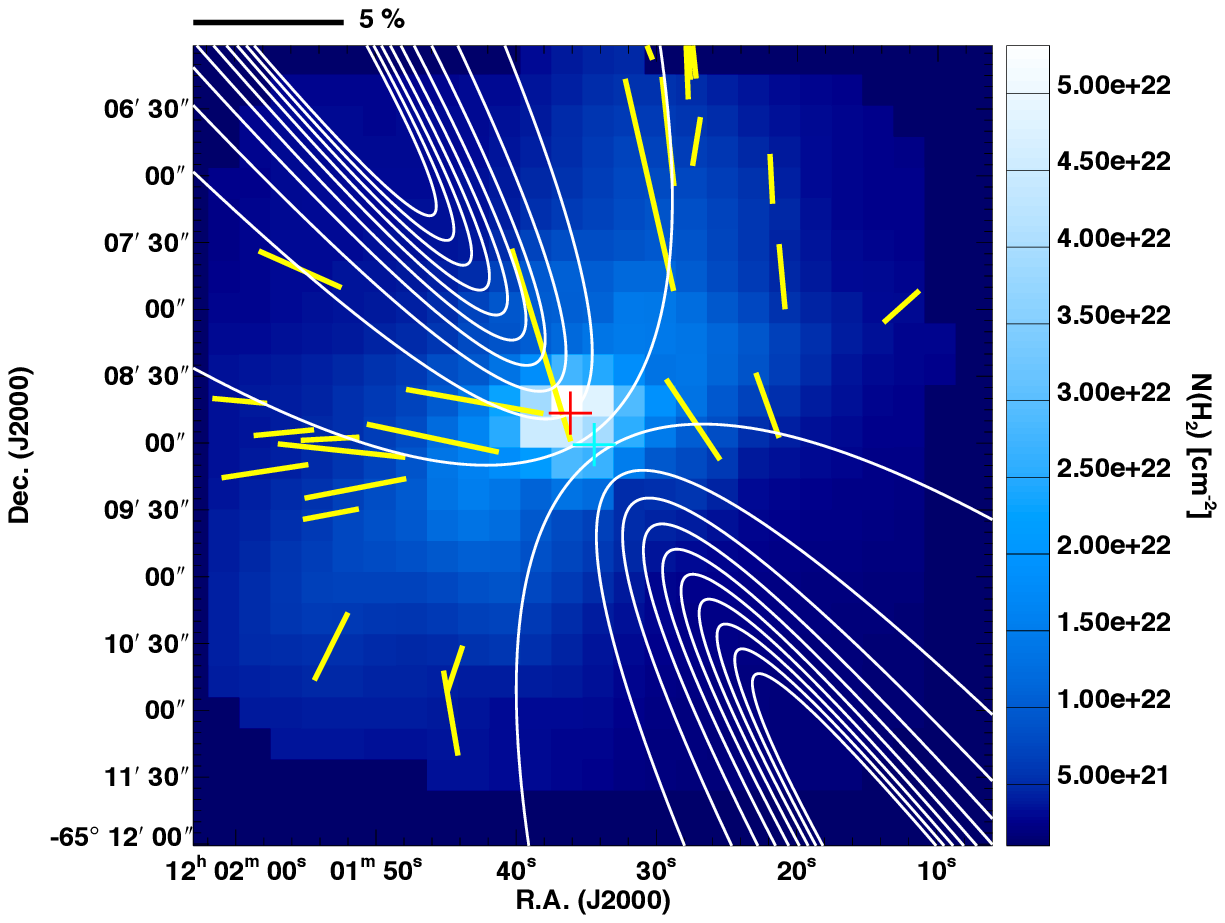}
\end{center}
 \caption{Same as Figure 8, but with the column density map used as the background image.}
   \label{fig}
\end{figure}

\clearpage 

\begin{figure}[t]
\begin{center}
 \includegraphics[width=6.5 in]{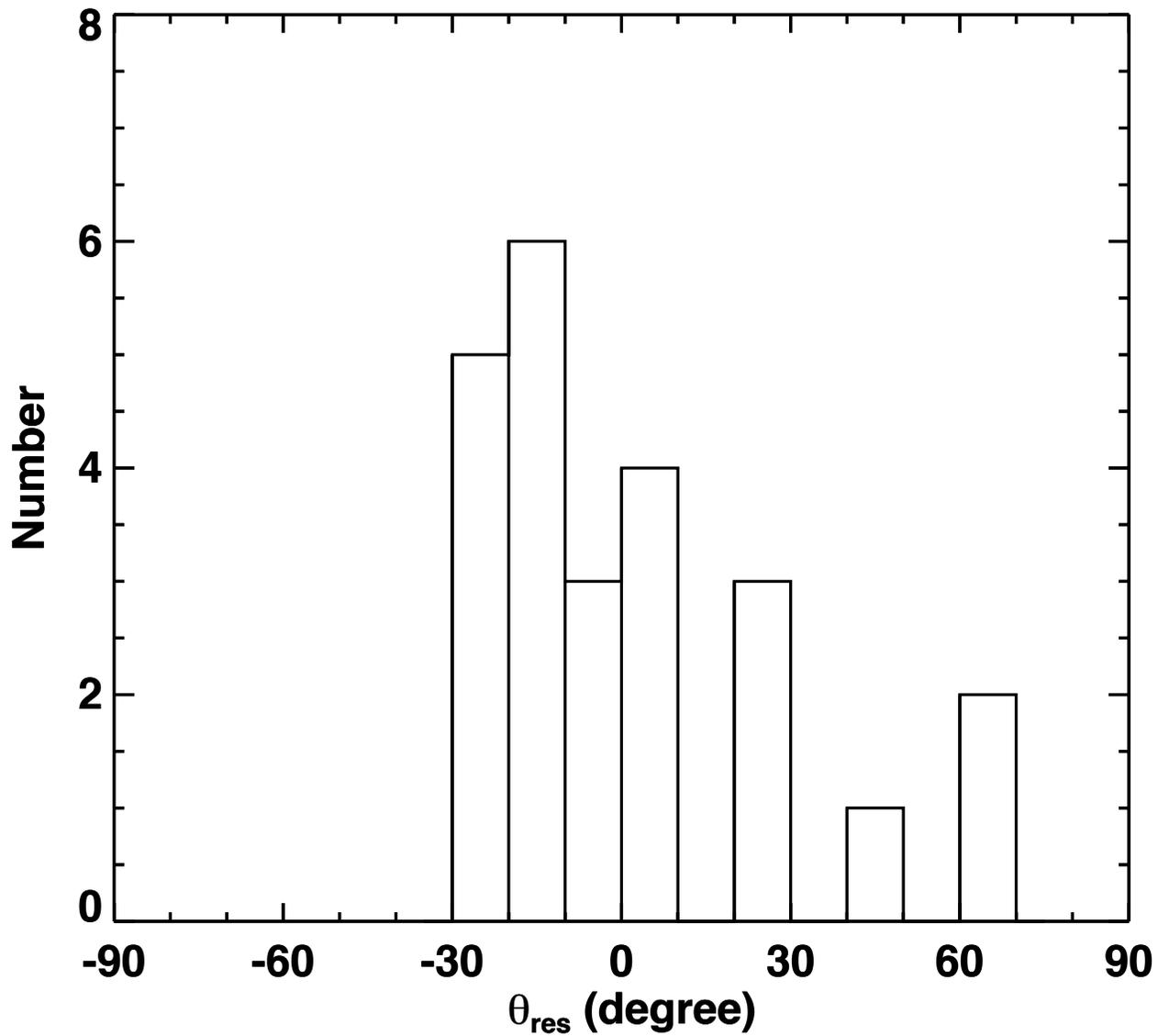}
\end{center}
 \caption{Histogram of residuals for the observed polarization angles after subtraction of the angles obtained by parabolic fitting ($\theta_{\rm res}$). All stars were used to make the histogram.}
   \label{fig}
\end{figure}

\clearpage 

\begin{figure}[t]
\begin{center}
 \includegraphics[width=6.5 in]{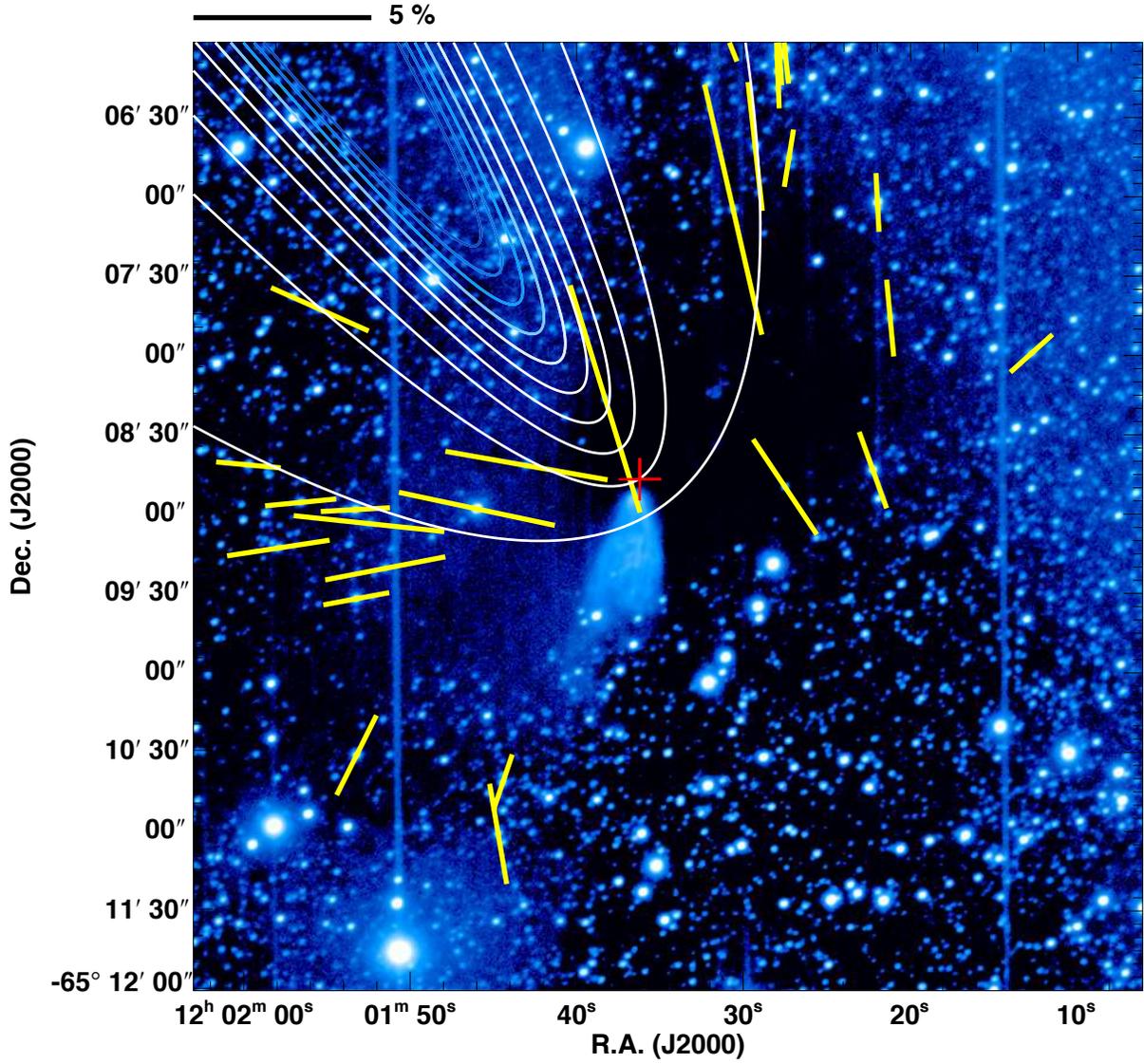}
\end{center}
 \caption{Polarization vectors after subtraction of the ambient components. Stars with $P_H / \delta P_H \ge 4$ and $P_H \ge 1.5\%$ are plotted. The white lines indicate the direction of the magnetic field inferred from parabolic fitting. The southern parabola was removed from the figure. The red plus sign shows the core center determined on the column density map based on {\it Herschel}. The scale bar above the image indicates 5\% polarization.}
   \label{fig}
\end{figure}

\clearpage 

\begin{figure}[t]
\begin{center}
 \includegraphics[width=6.5 in]{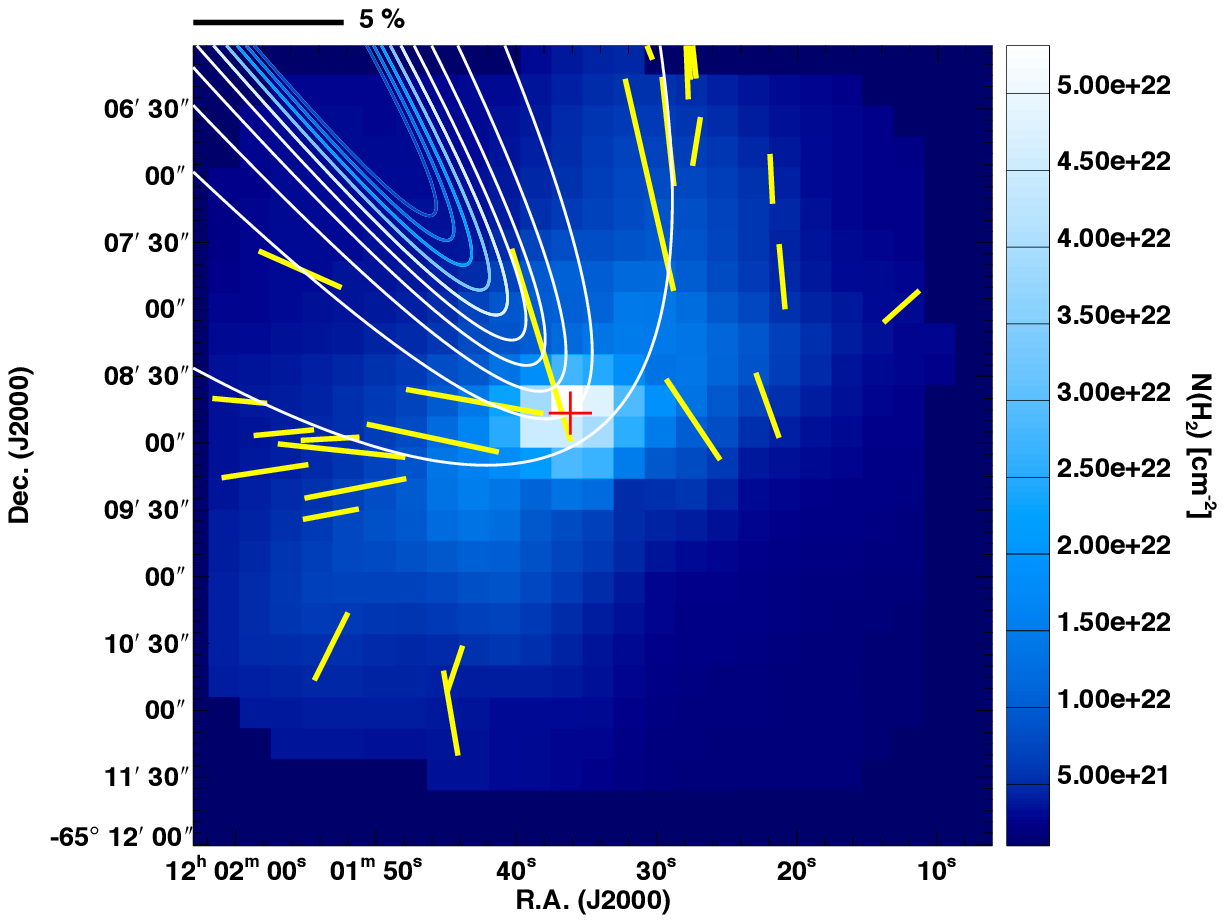}
\end{center}
 \caption{Same as Figure 11, but with the column density map used as the background image.}
   \label{fig}
\end{figure}

\end{document}